\documentclass[10pt,a4paper]{article}
\usepackage{amssymb,amsmath,graphicx,subcaption,hyperref,cite}

\usepackage[utf8]{inputenc}
\usepackage[english]{babel}

\begin{document}
\begin{center}
 \large{\bf Compensation in the spin-1/2 site diluted Ising ferrimagnet: A Monte Carlo study}
\end{center}

\vskip 1cm

\begin{center}{\it Sk Sajid$^{1}$ and Muktish Acharyya$^{2}$}\\
\vskip 0.2 cm
{\it Department of Physics, Presidency University,}\\
{\it 86/1 College street, Kolkata-700073, INDIA}\\
\vskip 1 cm
{1. E-mail: sajid.burd@gmail.com}\\
{2. E-mail: muktish.physics@presiuniv.ac.in}\end{center}

\begin{abstract}
A two-dimensional spin-1/2 trilayer magnetic system with quenched non-magnetic impurity is studied. The lattice is formed by alternate layers of two different theoretical atoms \textbf{A} and \textbf{B} arranged in a particular fashion \textbf{A-B-A}. The compensation point appears below the critical temperature, for which total magnetization of the system becomes zero even though the sublattice magnetization has a nonzero value. For a range of values of the relative interaction strength in the Hamiltonian, a compensation point is observed. We considered the Ising mechanics and employed the Monte Carlo method to determine the compensation point and critical temperature of the system. However, the effects of impurity in such systems are still not well studied. With that in mind, we address the effects of random non-magnetic impurity in the trilayer system. We also investigate the lattice morphologies in the presence of compensation and dilution and finally obtain the three-dimensional phase diagram for selected Hamiltonian parameters and impurity concentration.
\end{abstract}


\section{Introduction}
\label{intro}
Ferrimagnetic systems have been studied long experimentally and theoretically, due to their potential for  technological applications, such as the magneto-optical recording\cite{connell}, magnetocaloric effect\cite{phan} and giant magnetoresistance\cite{camley}. Besides these properties, one of the interesting phenomena of such layered materials is the presence of compensation temperature. Compensation point ($T_{comp}$) is a point below the critical temperature where the total magnetization becomes zero even though there exists a non zero sublattice magnetization. Under certain range of relative interaction strength, different temperature dependencies of different sublattices cause the appearance of the compensation point. It has been reported at this point some physical properties exhibit peculiar behaviour which includes diverging coercitivity\cite{connell,shieh}. Some ferrimagnetic materials are known to have a compensation point at room temperature and the fact that around this point coercive field ($H_c$) is strongly temperature dependent, makes it particularly useful for thermomagnetic recording devices\cite{connell}. With the experimental realization of the layered magnetic systems such as bilayer\cite{stier}, trilayer\cite{smits,leiner}, and multilayer\cite{kepa,chern,sankowski,chung,samburskaya}, theoretical studies are needed to provide a better understanding.\\
However, these complex systems do not have any theoretical solution except for a few handful of cases\cite{baxter}. Thus, approximation methods are required to tackle these systems. Trilayer spin-1/2 ferrimagnetic system is one of the simplest cases where compensation point has been observed\cite{branco17,branco18}. There have been efforts to study trilayer or three-layered superlattices via different approaches such as Monte Carlo simulation (MC)\cite{branco17,naji45} and mean-field approximation (MFA)\cite{naji399}. We also cite the study of extended ferrimagnetic structure analyzed within Monte Carlo method\cite{maaouni} \\
The effects of dilution (or disorder) to such materials is significant since the truly homogenous models are only an idealization and real magnets always have some sort of impurities present in them. Even though the single spin trilayer magnetic systems do not need any kind of dilution to show the compensation phenomenon unlike bilayers and multilayers, the question of whether the Neel temperatures remain same under dilution is not answered. Hence, it is interesting to see how the dilution affects the critical ($T_{crit}$) and the compensation temperature ($T_{comp}$) and thereby the phase boundary diagram. A Disordered layered system such as spin-1/2 and spin-1 diluted Ising bilayers and multilayers have been studied within various methods such as mean-field approximation (MC)\cite{kaneyoshi2} and Monte Carlo simulation (MC)\cite{branco16}.\\
Very recently, several important studies on layered magnetism have been found in the literature. 
The dilution effects on compensation temperature in nano trilayer graphene structure was 
studied \cite{fadil1} through the MC simulation. The compensation temperature has been found to increase
when the dilution probability decreases in ABA systems, whereas, for BAB system 
the compensation temperature decreases as the dilution probability is reduced. The magnetzation and
compensation behaviours in a mixed spin (7/2,1) antiferromagnetic ovalene nanostructures was studied by
MC approach. In this case, the appearance of two compensation temperatures is reported\cite{fadil2}.
The Monte Carlo studies are done in Blume Capel model of a bilayer graphyne structure with RKKY
interactions. It was observed\cite{fadil3} that the transition temperature increases with decreasing the number of
nonmagnetic layers. Although, there have not been many efforts to study the morphology and the phase diagram of a diluted trilayer material. Therefore, we conduct this study on a trilayer spin-1/2 lattice via Monte Carlo (MC) approach. In Sec. 2 we introduce our trilayer model and write down its Ising Hamiltonian. We discuss the simulation and data analysis techniques in Sec. 3. We next, show our results in Sec. 4, and finally, we present our conclusion and remarks in Sec. 5.  
\newpage

\section{The Model}
\label{model}
\begin{figure}[h]
\begin{center}
\resizebox{8cm}{!}{\includegraphics{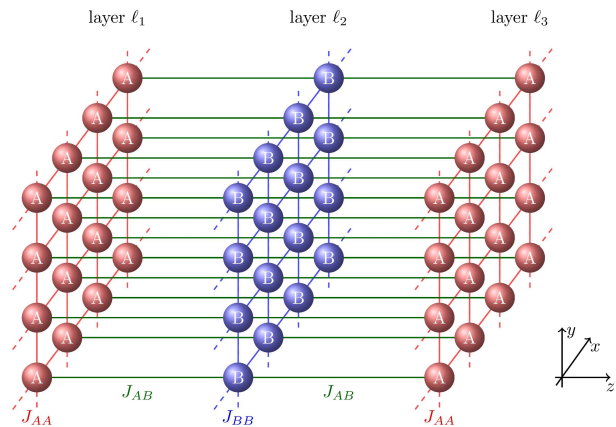}}
\caption{ Schematic representation of ABA trilayer system in which $J_{11}=J_{33}=J_{AA}>0; 
J_{12}=J_{23}=J_{AB}<0; J_{22}=J_{BB}>0$. This image is taken from the reference\cite{branco17}.}
\label{trifig}
\end{center}
\end{figure}

We study the trilayer magnetic system (as shown in Fig-\ref{trifig}) consisting of three monoatomic layers $l_{1},l_{2},l_{3}$ each of which is composed exclusively of either type-A or type-B atoms. The general system is described by the spin-1/2 Ising Hamiltonian without an external field,\\
\begin{align}\label{eq:hamiltonian}
\mathcal{H}=-
 J_{11}\sum_{\langle ij\rangle}s_i s_j 
-J_{22}\sum_{\langle ij\rangle}s_i s_j
-J_{33}\sum_{\langle ij\rangle}s_i s_j
-J_{12}\sum_{\langle ij\rangle}s_i s_j
-J_{23}\sum_{\langle ij\rangle}s_i s_j,
\end{align}\\
where $\langle ij\rangle$ indicate summations over all distinct pairs of nearest-neighbour sites in the same layer, whereas $ij$ are over pairs of nearest-neighbour sites in adjacent layers. The spin variables $s_n$ can take the value "+1", "-1". The exchange parameters are  $J_{AA}>0$ for A-A bonds, $J_{BB}>0$ for B-B bonds, and $J_{AB}<0$ for A-B bonds. There exists a periodic boundary condition (PBC) in both the x, y direction (intraplanar) where the z plane (interplanar) has the free boundary condition. To mimic the impurity we have used s=0 spin value at the site of impurity which should not to be confused with a spin-1 system. There are two possible configurations of the trilayer with more type-A atoms than type-B, e.g., \textbf{AAB} and \textbf{ABA}, but in this work, we consider only one type of configuration, i.e., \textbf{ABA}. \textbf{ABA} system corresponds to $J_{11}=J_{33}=J_{AA}, J_{12}=J_{23}=J_{AB}, J_{22}=J_{BB}$. Note that in the scope of this study the Boltzmann constant, $k_B = 1$ for the sake of simplicity.

\section{Monte Carlo simulation}
\label{mcsim}
The model described before was simulated using the standard importance sampling technique. For the Monte Carlo simulation, we employed the Metropolis algorithm ($w(s_i\rightarrow s_f) = min [e^{-\frac{\Delta E}{k_BT}},1]$) to analyze the Hamiltonian on a three stacked square lattice with $L^2$ sites each. The lattice size is taken to be $100\times100$. As shown in Ref-\cite{branco17}, for $L>60$ compensation point tends to a stable value. Since the study of the critical exponents is not within the scope of this literature, the lattice size is quite standard. The whole system starts from a higher temperature and then cooled down to a lower temperature. Hence, the initial spin configurations were taken either "+1" or "-1" randomly with equal probabilities. After initial spins are oriented, we put fixed non-magnetic impurities randomly with certain concentration at different sites in each layer. We flip the spins according to the Metropolis acceptance condition\cite{binder}. In each MC step, we take $L^2$ number of trials to flip the spins. Here in this study, we take a $100\times100$ lattice and perform $50,000$ MC steps in a single simulation. We discarded up to $10,000$ steps to account for the equilibrium.
Our algorithm then calculates sublattice magnetization, defined as, 
\begin{equation}
M_{\sigma}= \frac{1}{N} \sum_i{s_i}
\end{equation}
where $N=L^2$ is the number of sites in $\sigma$ layer and total magnetization is,
\begin{equation}
M_{tot}= \frac{1}{3} (M_1+M_2+M_3)
\end{equation}
At $T=T_{comp}$ sublattice magnetizations $M_\sigma$ cross each other and the total magnetization $M_{tot}$  becomes zero. Then due to symmetry
\begin{equation}
\vert M_2\vert=\vert M_1+M_3\vert\label{eq:1}
\end{equation} 
and
\begin{equation}
sgn(M_1)=-sgn(M_2) ; sgn(M_3)=-sgn(M_2)\label{eq:2}
\end{equation}
As the middle layer B is influenced with antiferromagnetic interaction by both the A layers, B-layer magnetization $M_2$ always gets saturated  in the opposite direction with respect to both the A layers.
At equlibrium, we calculate thermodynamic quantities like susceptibility $\chi$ and specific heat $C_v$ as follows:
\begin{equation}
\chi_\sigma = N_\sigma \frac{(\langle M^2_\sigma\rangle-\langle M_\sigma\rangle ^2)}{k_B T}
\end{equation} 
and
\begin{equation}
(C_v)\sigma = N_\sigma \frac{(\langle E^2_\sigma\rangle-\langle E_\sigma\rangle ^2)}{k_B T^2}
\end{equation}
where $\sigma=1,2,3$ and $N_1=N_2=N_3=L^2$ is the number of sites present in each sublattice. Therefore as the equations \eqref{eq:1}, \eqref{eq:2} suggests both $T_{crit}$ and $T_{comp}$ can be calculated by looking at the total magnetization curve as $M_{tot}$ vanishes at those points. To obtain $T_{crit}$ and $T_{comp}$ the simulation was performed in the vicinity of the critical point and the compensation point. We divide this range up to 50 temperature points considering a reasonable amount of simulation time otherwise the equilibrium would take too long. For this number of points in a short space, the interval between two temperature points can be assumed to be a straight line and thus we employed linear interpolation between points where the $sgn(M_{tot})$ changes in order to obtain the said points. The errors are given by linear interpolation error bound \cite{scarborough}. These obtained temperature points were also verified by the susceptibility ($\chi$) and specific heat ($C_v$) curves. Next, for the phase boundary curve, we calculated $T_{crit}$ and $T_{comp}$ at different dilution while keeping the coupling parameters fixed, we varied the coupling parameters and obtained both $T_{crit}$ and $T_{comp}$ for various impurity concentration. Thus we calculated $T_{crit}$ and $T_{comp}$ for different concentration of the site dilution as well as for different Hamiltonian parameters and a three-dimensional phase diagram was obtained.

\section{Results}
\label{res}
We have studied the thermodynamic and magnetic response of a trilayer system along with its morphology in the presence of non-magnetic impurity with the help of MC simulation. Our goal is to observe the effects of Hamiltonian parameters and impurity concentration on the critical temperature and the compensation temperature and finally obtain a phase boundary for both of these.\\ 
In the case of the study of morphology, we first obtained the results for the pure system in the presence of compensation effect.
\subsection{Lattice Morphology}
\begin{figure}[h!]
  \begin{subfigure}[b]{0.3\textwidth}
    \includegraphics[width=\textwidth]{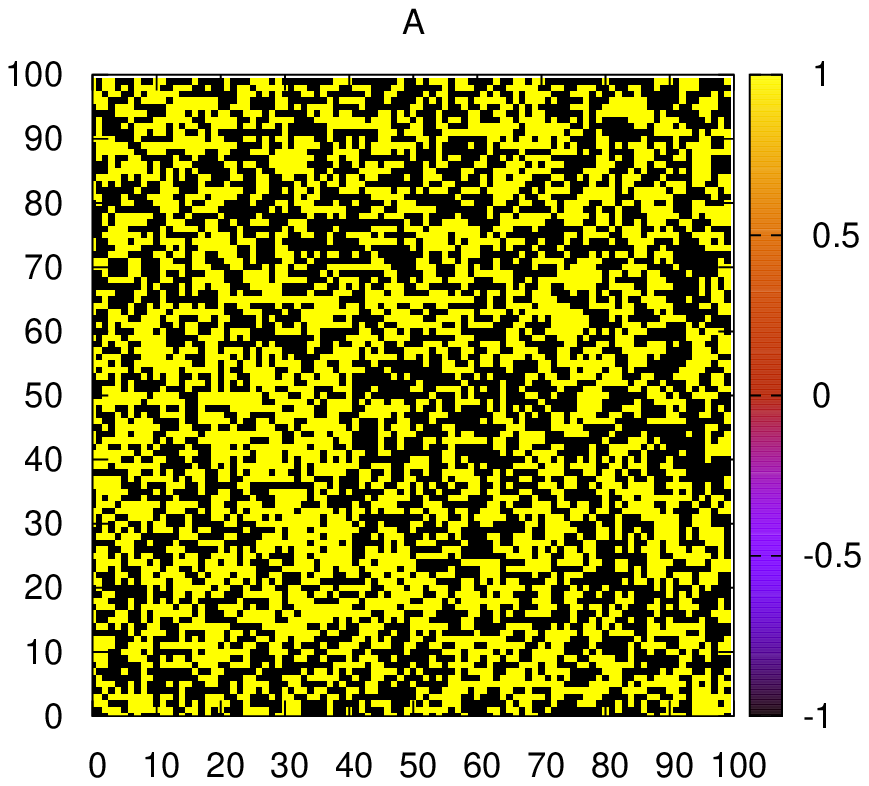}
  \end{subfigure}
  \begin{subfigure}[b]{0.3\textwidth}
    \includegraphics[width=\textwidth]{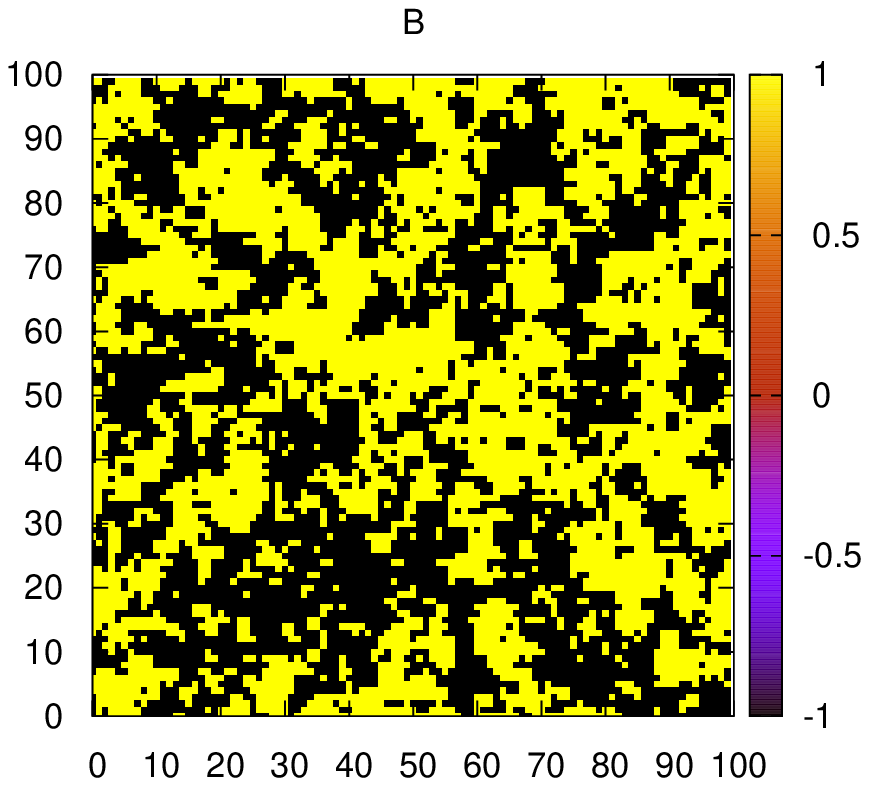}
  \end{subfigure}
  \begin{subfigure}[b]{0.3\textwidth}
    \includegraphics[width=\textwidth]{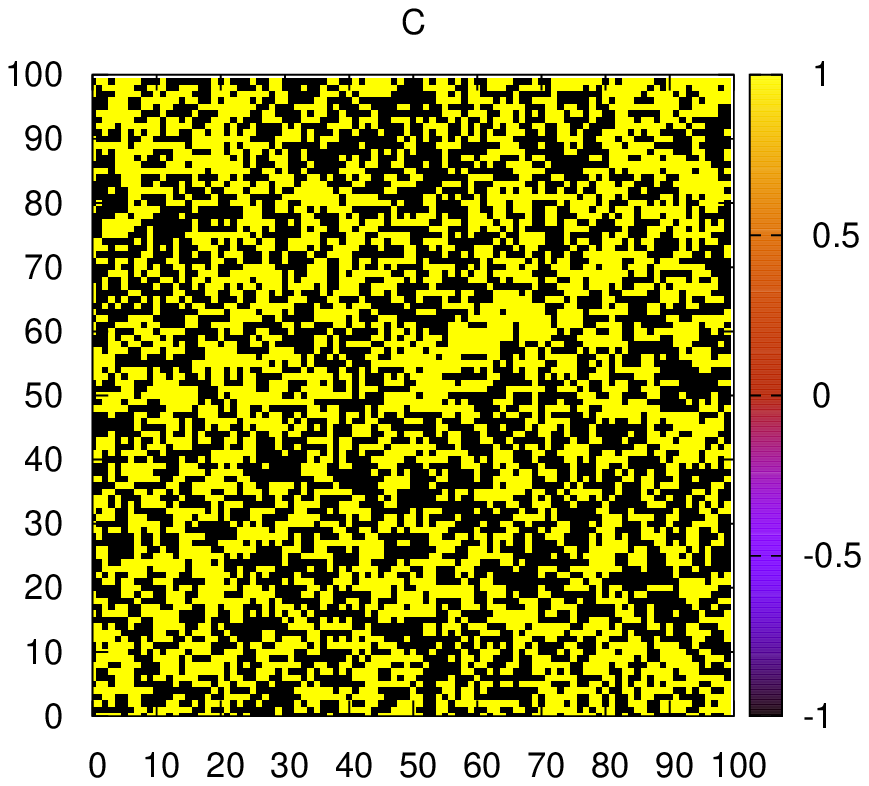}
  \end{subfigure}
  \caption{Density map of the spin matrix in the presence of the compensation ($J_{AA}/J_{BB}=0.5 , J_{AB}/J_{BB}=-0.1$) at $T= T_{crit}$ for  A-B-A respectively. The sublattice magnetizations are: $M_A=-5.29\times10^{-3}, M_B=1.70\times10^{-2}, M_C=-5.11\times10^{-3}$}
\label{fig:morph1crit}
\end{figure}

\begin{figure}[h!]
  \begin{subfigure}[b]{0.3\textwidth}
    \includegraphics[width=\textwidth]{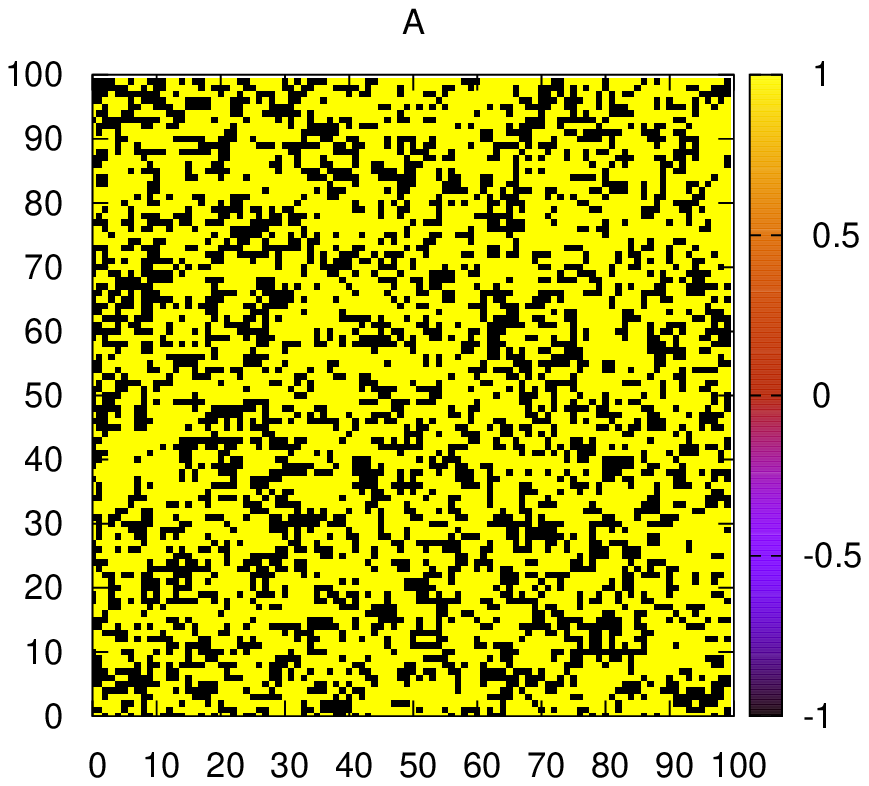}
  \end{subfigure}
  \begin{subfigure}[b]{0.3\textwidth}
    \includegraphics[width=\textwidth]{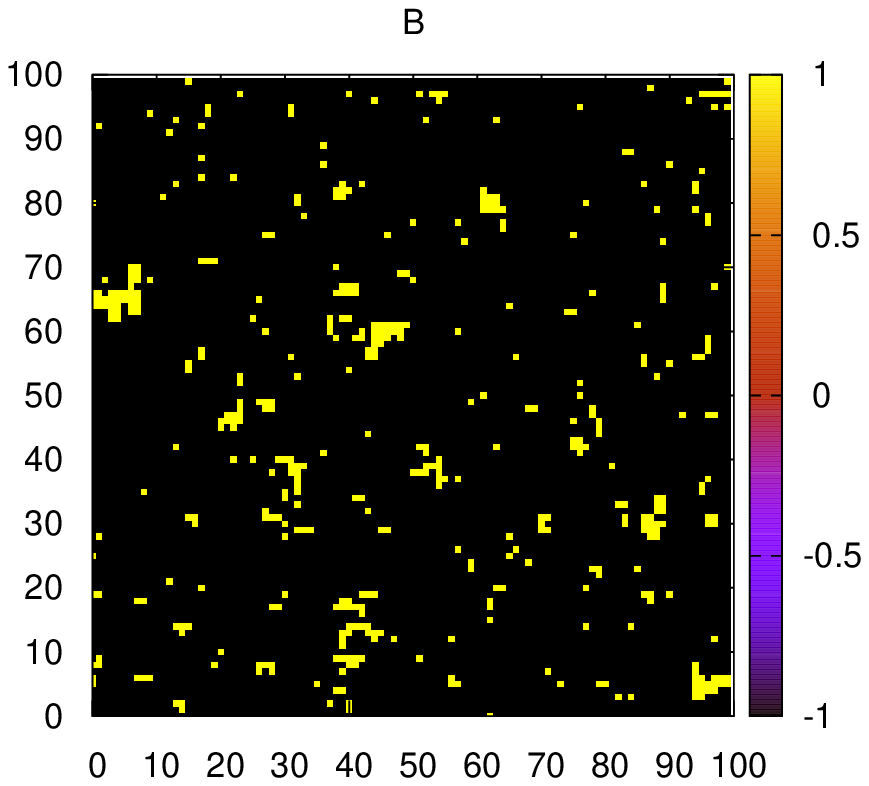}
  \end{subfigure}
  \begin{subfigure}[b]{0.3\textwidth}
    \includegraphics[width=\textwidth]{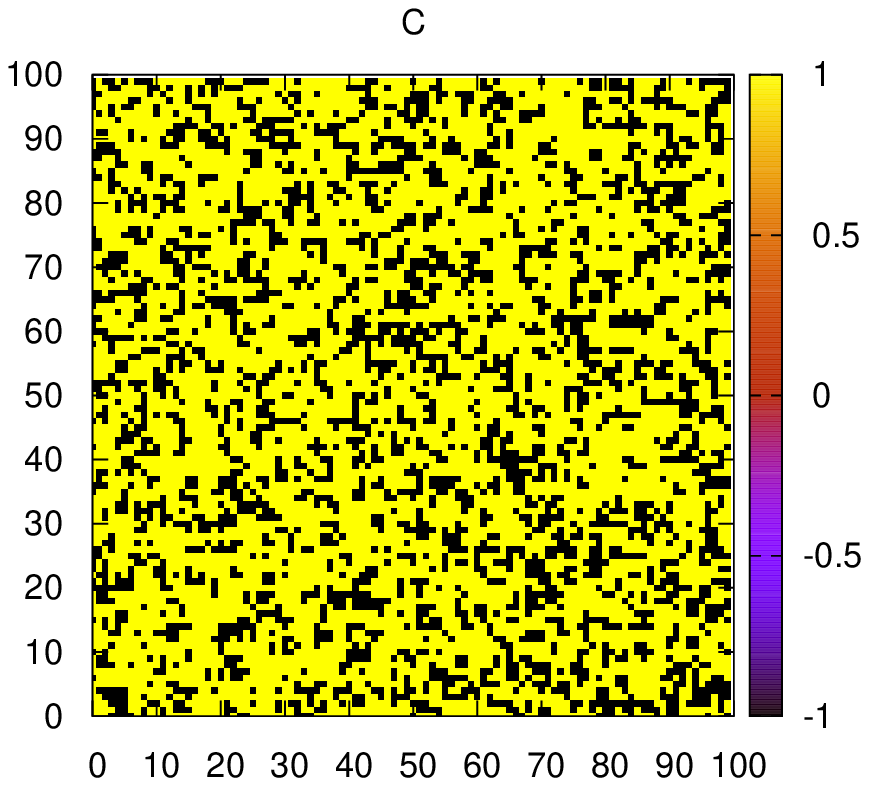}
  \end{subfigure}
  \caption{Density map of the spin matrix in the presence of the compensation ($J_{AA}/J_{BB}=0.5 , J_{AB}/J_{BB}=-0.1$) at $T= T_{comp}$ for A-B-A respectively. The sublattice magnetizations are: $M_A=0.49, M_B=-0.96, M_C=0.49$}
\label{morph1comp}
\end{figure}

As shown in Ref-\cite{branco17}, we selected interaction  strength ($J_{AA}/J_{BB}=0.5 , J_{AB}/J_{BB}=-0.1$) in the Figs-\ref{fig:morph1crit},\ref{morph1comp} such that the compensation effect is present. The yellow dots denote up spins while the black dots represent down spins. Fig-\ref{fig:morph1crit} and Fig-\ref{morph1comp} are density maps at $T_{crit}$ and $T_{comp}$ respectively. The fact that $T_{crit}$ and $T_{comp}$ are two different points by nature is reinforced by their lattice morphology. It is evident that Fig-\ref{fig:morph1crit} and Fig-\ref{morph1comp} are morphologically different. Every layer is occupied by almost an equal amount of up and down spins. Hence the sublattice magnetizations at $T_{crit}$ (Fig-\ref{fig:morph1crit}) are practically vanishing. However, it is interesting to note the value of the sublattice magnetizations at this critical temperature. We have magnetizations in the order of $10^{-3}$ in both the A layers while on the other hand, the B layer has the magnetization in the order of $10^{-2}$. This can be understood from the morphology at $T_{crit}$ as shown in Fig-\ref{fig:morph1crit}. Since the B layer has relatively larger spin clusters compared to the A layers, the value of the sublattice magnetization in the mid-layer is higher than the rest.\\  
Now in the case of $T_{comp}$ (Fig-\ref{morph1comp}), sublattices are dominated by either of the spins. For instance, both the A layers is dominated by up spins where the B layer is mostly occupied by down spins. This leads to non-vanishing layered magnetizations at $T_{comp}$. Similar to the case at $T_{crit}$, the difference in the size of the spin clusters, creates an asymmetry in the layered magnetization. As a result, the total magnetization becomes zero and we see the signature of the compensation in the Fig-\ref{morph1comp}.\\

\begin{figure}[h!]
  \begin{subfigure}[b]{0.3\textwidth}
    \includegraphics[width=\textwidth]{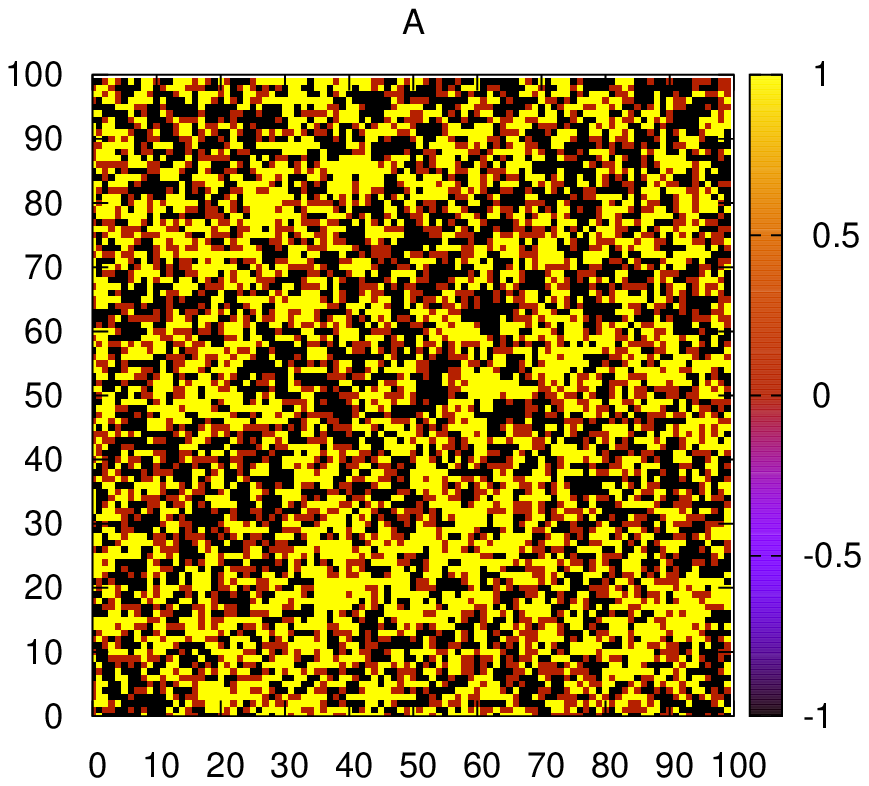}
  \end{subfigure}
  \begin{subfigure}[b]{0.3\textwidth}
    \includegraphics[width=\textwidth]{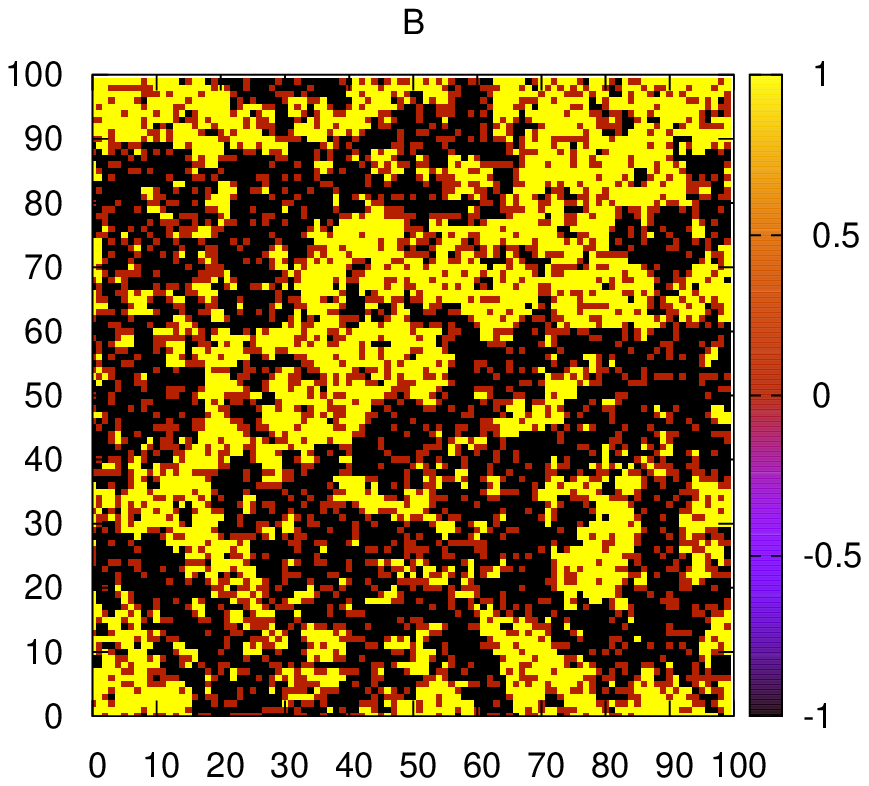}
  \end{subfigure}
  \begin{subfigure}[b]{0.3\textwidth}
    \includegraphics[width=\textwidth]{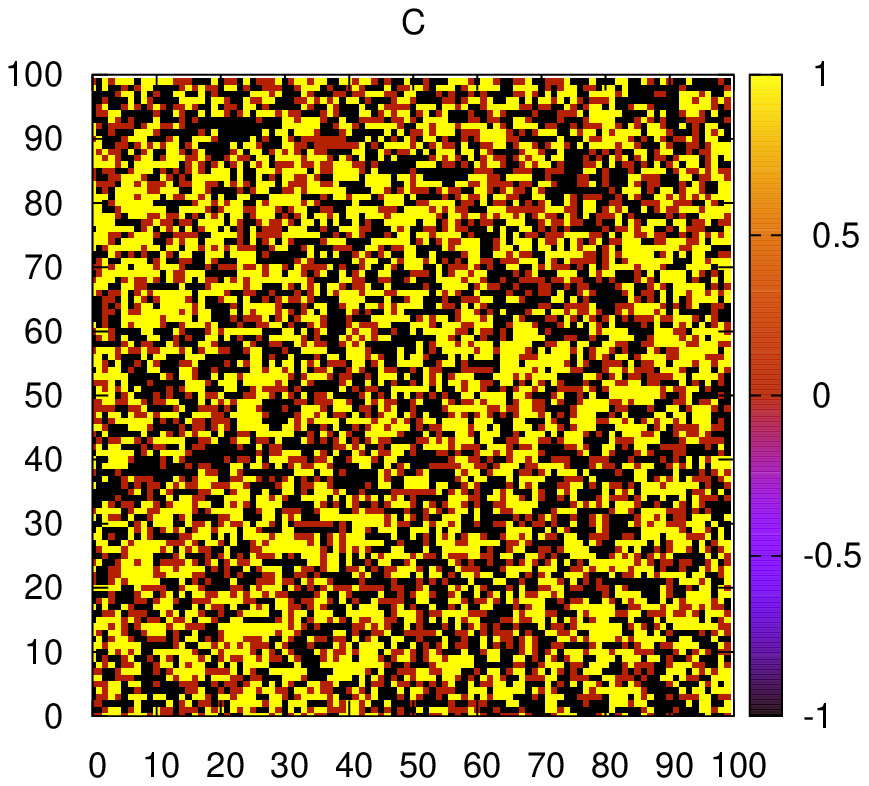}
  \end{subfigure}
  \caption{Latttice morphology of the diluted system ($C=0.26$) in the presence of the compensation ($J_{AA}/J_{BB}=0.5 , J_{AB}/J_{BB}=-0.1$) at $T=T_{crit}$ for A-B-A respectively. The sublattice magnetizations are: $M_A=3.25\times10^{-3}, M_B=1.72\times10^{-2}, M_C=-3.09\times10^{-3}$}
\label{fig:morph2crit}
\end{figure}

\begin{figure}[h!]
  \begin{subfigure}[b]{0.3\textwidth}
    \includegraphics[width=\textwidth]{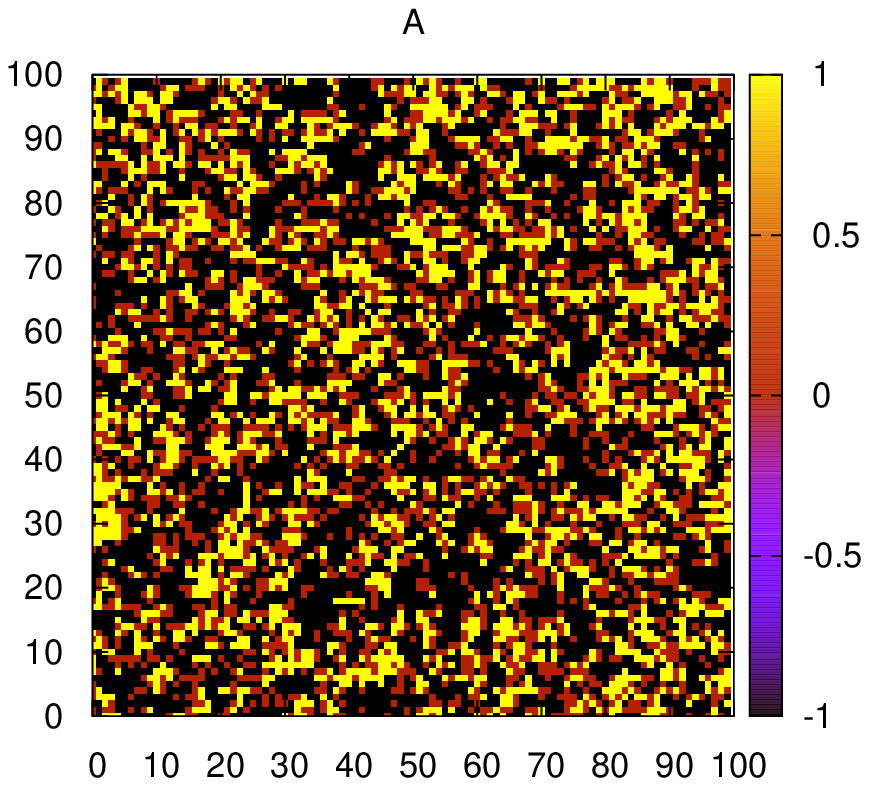}
  \end{subfigure}
  \begin{subfigure}[b]{0.3\textwidth}
    \includegraphics[width=\textwidth]{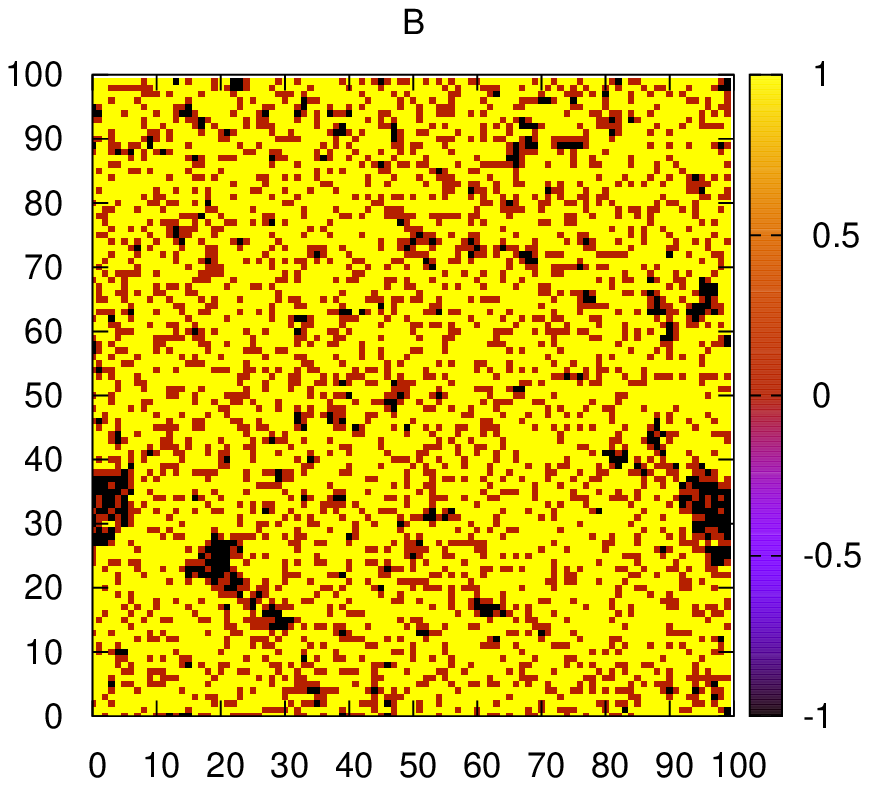}
  \end{subfigure}
  \begin{subfigure}[b]{0.3\textwidth}
    \includegraphics[width=\textwidth]{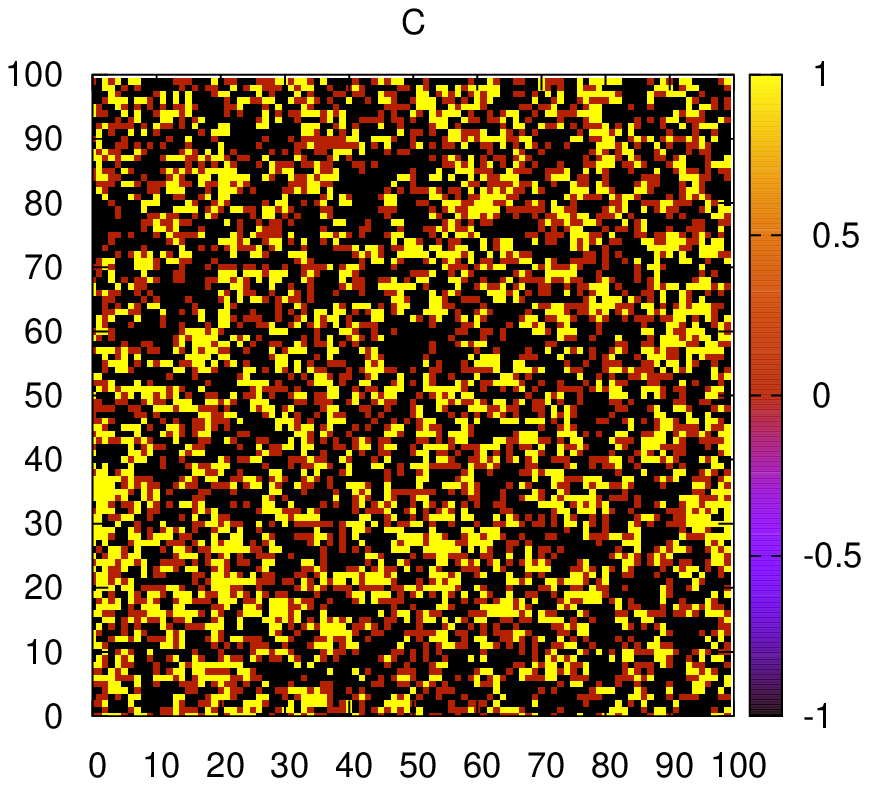}
  \end{subfigure}
  \caption{Latttice morphology of the diluted system ($C=0.26$) in the presence of the compensation ($J_{AA}/J_{BB}=0.5 , J_{AB}/J_{BB}=-0.1$) at $T=T_{comp}$ for A-B-A respectively. The sublattice magnetizations are: $M_A=-0.36, M_B=0.70, M_C=-0.35$}
\label{fig:morph2comp}
\end{figure}

Finally, we introduce certain impurity in the system ($C=0.26$) without changing the exchange parameters ($J_{AA}/J_{BB}=0.5 , J_{AB}/J_{BB}=-0.1$) and observe the lattice morphology that is shown in Figs-\ref{fig:morph2crit}, \ref{fig:morph2comp}. The brown dots in the density map represents the impure points in the spin matrix or s=0. We report that the lattice morphology at $T_{crit}$ and $T_{comp}$ remains unchanged after adding site dilution except for the introduction of impure points (brown dots). Although in this case the sublattice magnetization at $T_{comp}$ (Fig-\ref{fig:morph2comp}) is dominated by opposite spins compared to the case of without impurity. 
\newpage
\subsection{Thermal and Magnetic properties}
\begin{figure}[h]
\begin{center}
	\resizebox{8cm}{!}{\includegraphics{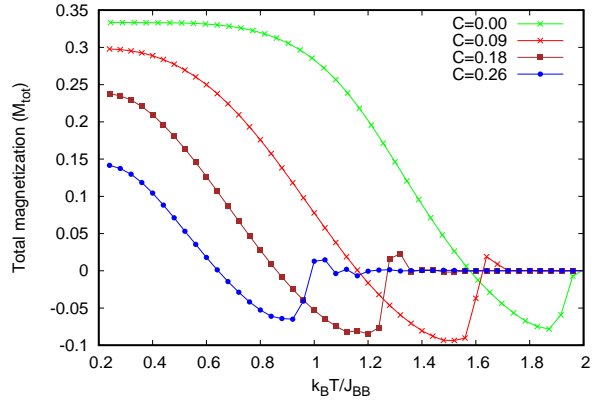}}\\
\caption{Magnetization(M) as a function of dimensionless temperature ($k_BT/J_{BB}$) for different dilutions, $C=0.0, 0.09, 0.18, 0.26$  }
\label{fig:mag}
\end{center}
\end{figure}

\begin{figure}[h]
\begin{center}
	\resizebox{8cm}{!}{\includegraphics{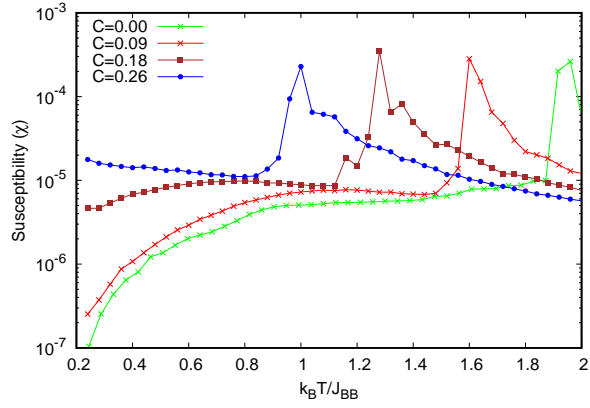}}
\caption{Semilog plot of susceptibility($\chi$) as a function of dimensionless temperature ($k_BT/J_{BB}$) for different dilutions, $C=0.0, 0.09, 0.18, 0.26$}
\label{fig:suscep}
\end{center}
\end{figure}

\begin{figure}[h]
\begin{center}
	\resizebox{8cm}{!}{\includegraphics{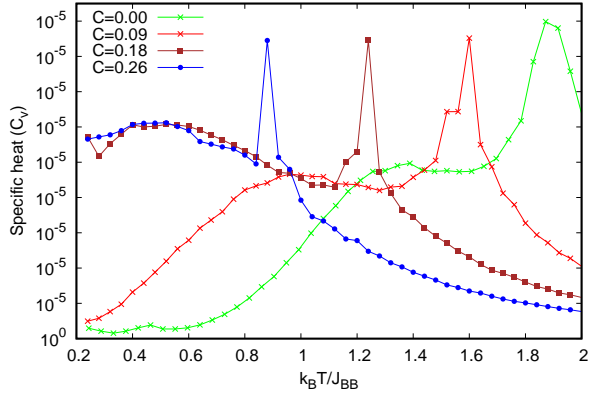}}\\
\caption{Specific heat($C_v$) as a function of dimensionless temperature ($k_BT/J_{BB}$) for different dilutions, $C=0.0, 0.09, 0.18, 0.26$}
\label{fig:cv}
\end{center}
\end{figure}

\begin{figure}[h!]
\begin{center}
\resizebox{8.0cm}{!}{\includegraphics[angle=0]{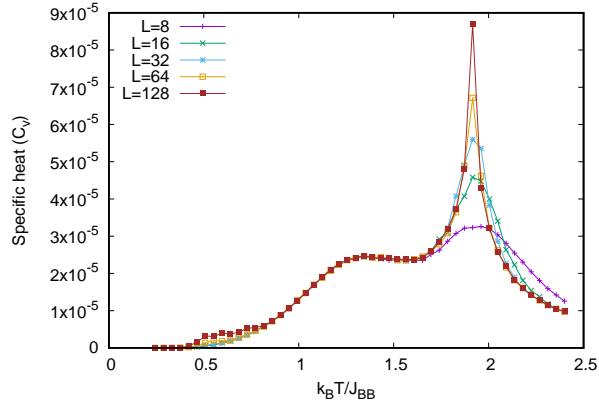}}
\caption{Specific heat ($C_v$ ) as a function of dimensionless temperature ($k_BT/J_{BB}$) for different lattice sizes L ranging from 8 to 128.}
\label{fig:fsscv}
\end{center}
\end{figure}

\begin{figure}[h!]
\begin{center}
\resizebox{8.0cm}{!}{\includegraphics[angle=0]{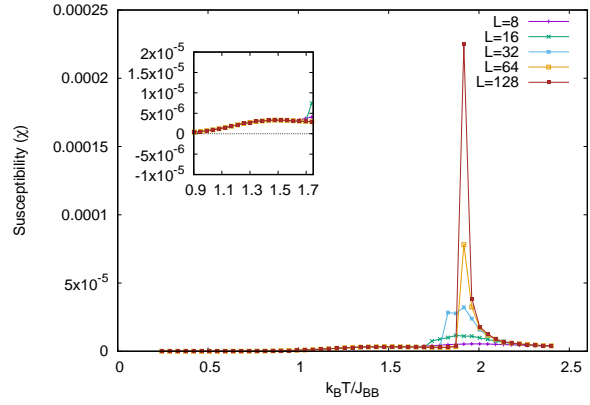}}
\caption{Susceptibility ($\chi$) as a function of dimensionless temperature ($k_BT/J_{BB}$) for different lattice sizes L ranging from 8 to 128.}
\label{fig:fsschi}
\end{center}
\end{figure}

In the previous section, we have found the presence of compensation in the pure and diluted system respectively. In order to confirm the result, we start analyzing by comparing the magnetization curve under compensation at different dilution concentration($C$). Fig-\ref{fig:mag} shows the total magnetization ($M_{tot}$) curve of the system for $C=0.00, 0.09, 0.18, 0.26$ respectively. In the figure, the thermal variation of the total magnetization $M_{tot}$ curve shows a clear sign of compensation phenomenon as predicted in the lattice morphologies. At $T\rightarrow0$ the spins of both the A-layers takes the value of $s=+1$ ($M_A=+1$) state. On the other hand, B-layer spins are dominated by $s=-1$ spins resulting $M_B=-1$. With the increase of thermal excitation layered magnetizations decreases from its saturated value. At $T_{comp}$ two positive sublattice magnetization of A-layer is such that it cancels out the negative sublattice	 magnetization of B-layer and remains positive while at $T_{crit}$ sublattice magnetization as well as total magnetization reduces to zero. {\it It can be seen that both $T_{crit}$ and $T_{comp}$ decreases with increasing dilution effect}. This can be understood as the magnetization is a measure of net magnetic moment per unit volume, with increasing impurity the number of active sites decreases and so does the net magnetic moment. It is worth noting that the saturation magnetization decreases as the dilution increases since we normalized the magnetization by the total number of sites and not by the active site present in the system. We see in Figs.-\ref{fig:suscep}, \ref{fig:cv} the susceptibility ($\chi$) and the specific heat ($C_v$) curve as a function of the dimensionless temperature $T$. It is evident that with increasing dilution the critical point and the compensation point shifts towards a lower temperature confirming the effects we see in the $M_{tot}$ vs $T$ curve.\\
\newpage
An examination of the finite size effect is done by plotting the thermodynamic quantities, such as specific heat and susceptibility. This reveals, as shown in Fig-\ref{fig:fsscv}, the high-temperature peak or $T_{crit}$ shows singular behaviour as $L\rightarrow\infty$ in both of these curves whereas the low-temperature peak or $T_{comp}$ does not show any of such singularity instead we get a plateau type of region. This result is very much similar to the case of the bilayers where low-temperature peak essentially remains unchanged for different $L$ and thus indicates the alteration of the short-range order\cite{landau,kim}. Very much similar picture is seen in the susceptibility curve (see Fig-\ref{fig:fsschi}) as well, however, we report a slight shift of the high-temperature peak positions for different system sizes. Since the low-temperature region is overwhelmed by the high-temperature peak in the Fig-\ref{fig:fsschi}, an inset of the plateau is attached for easier comprehension.
\newpage
\subsection{Three dimensional phase diagram}
\begin{figure}[h!]
\begin{center}
\begin{tabular}{c}
\resizebox{8.0cm}{!}{\includegraphics[angle=0]{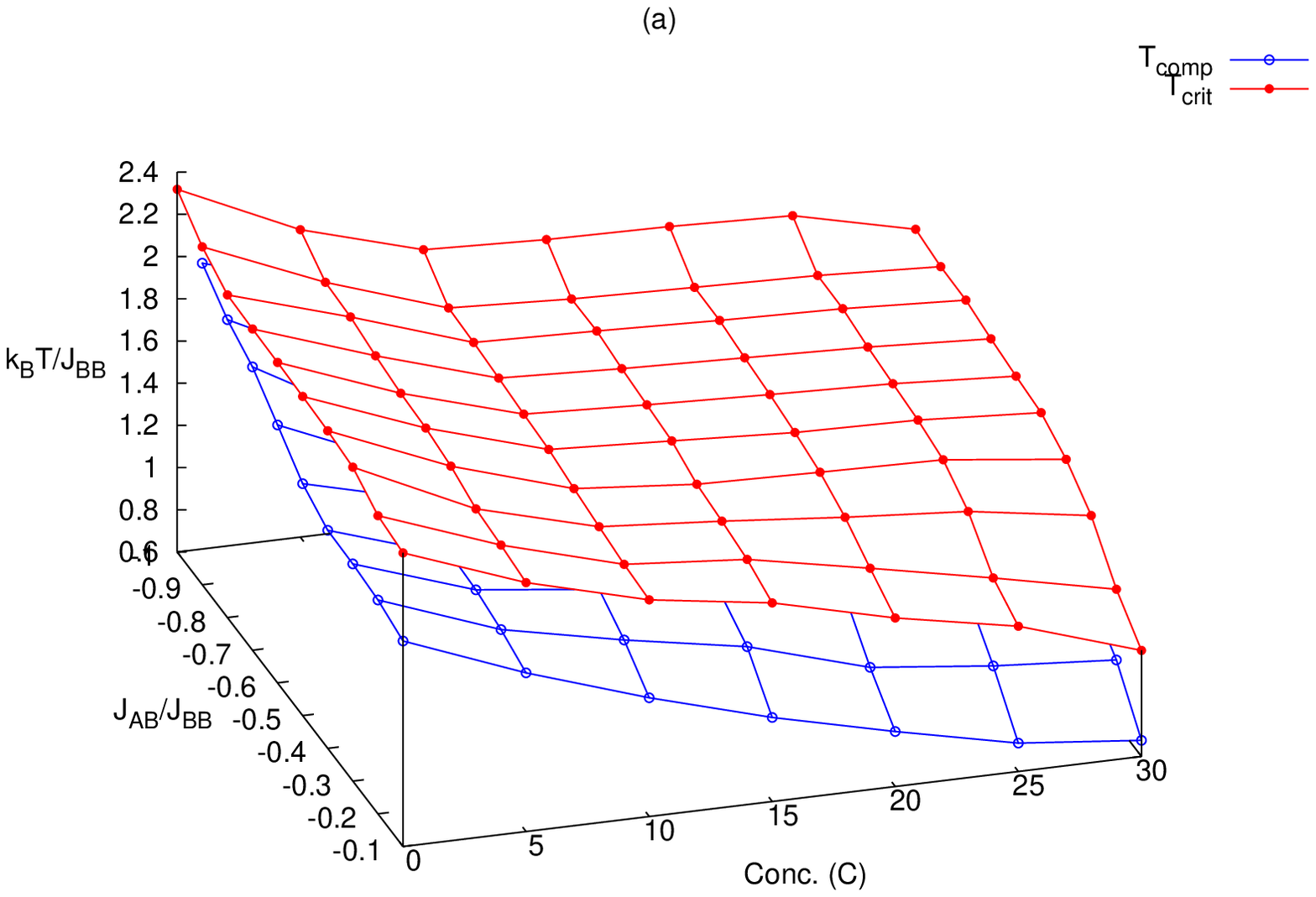}}
\resizebox{8.0cm}{!}{\includegraphics[angle=0]{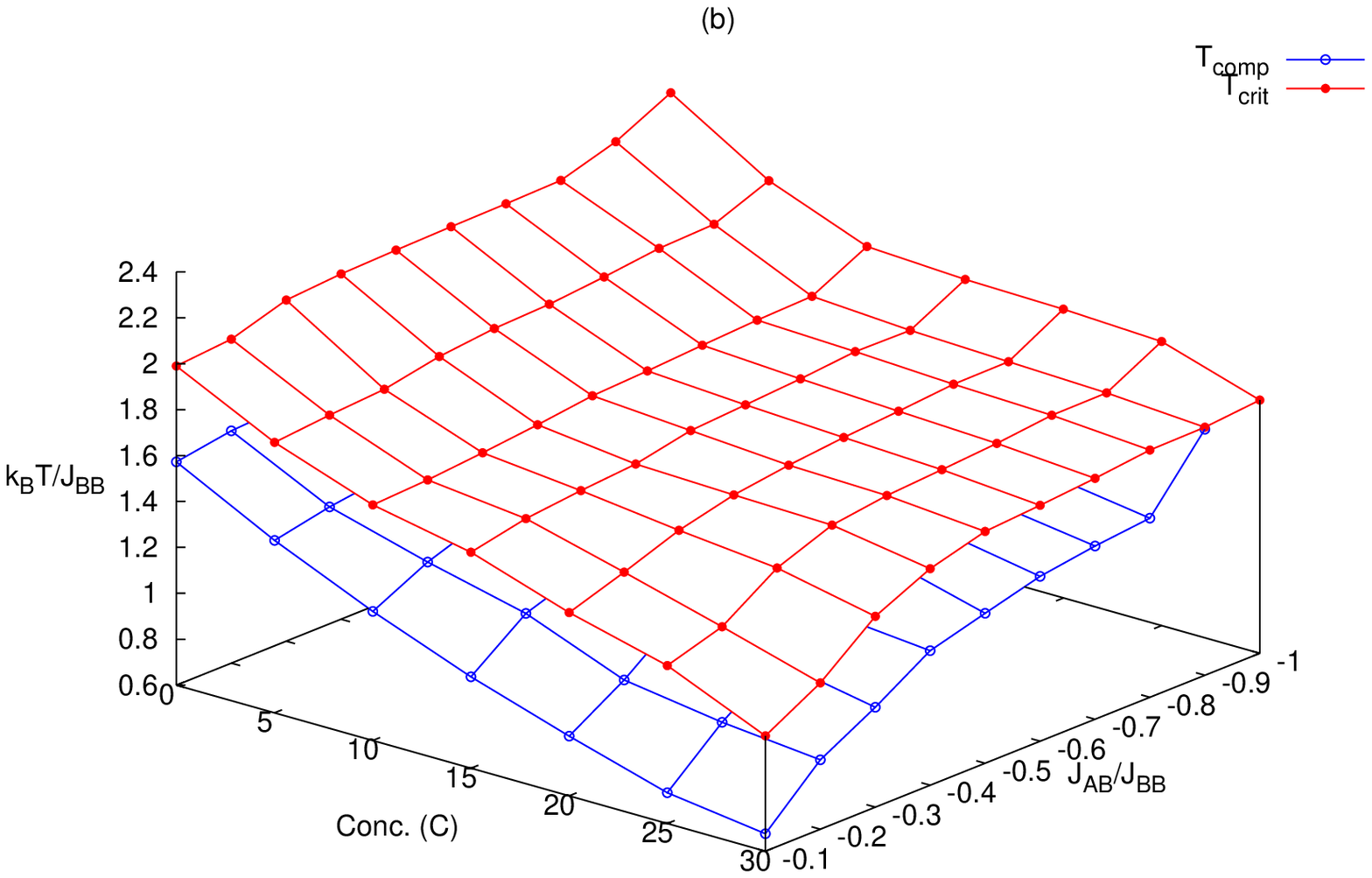}}
\\
\\
\resizebox{8.0cm}{!}{\includegraphics[angle=0]{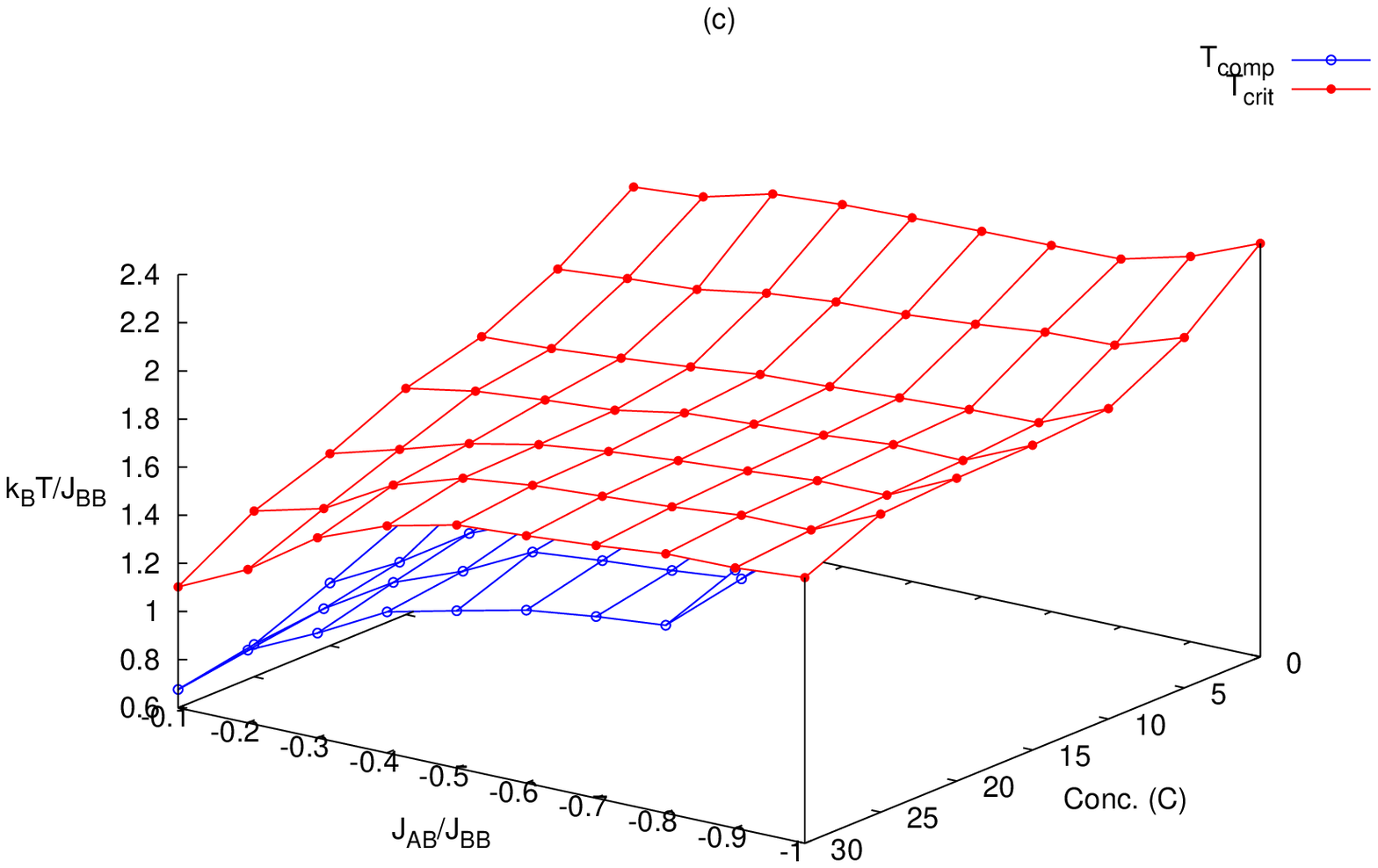}}
\resizebox{8.0cm}{!}{\includegraphics[angle=0]{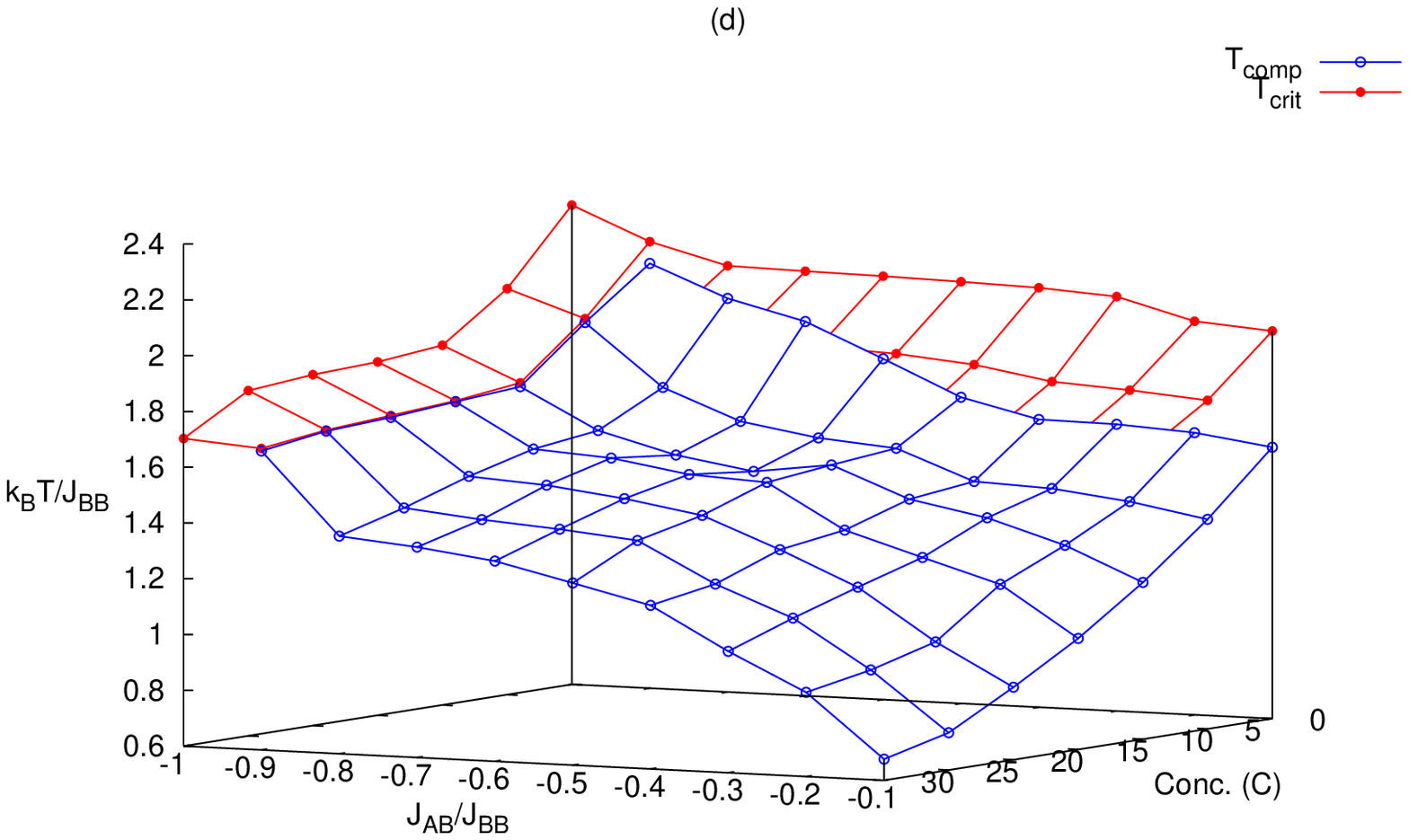}}
          \end{tabular}
\caption{Dimensionless critical temperature $k_BT_{crit}/J_{BB}$ (solid points) and dimensionless compensation temperature $k_BT_{comp}/J_{BB}$ (hollow points) as a function of dilution ($C$) and antiferromagnetic interaction strenth ($J_{AB}/J_{BB}$). The solid dots denote the critical points ($T_{crit}$) and the hollow dots represent compensation points ($T_{comp}$). The error bars are smaller than the symbols used. The surface bifurcates around $J_{AB}=-0.9$ which means above this point there is no compensation. Different angels of the 3d plot is shown for easier comprehension.}
\label{fig:pbd}
\end{center}
\end{figure}

In the Fig-\ref{fig:pbd} the effect of dilution and Hamiltonian parameter on both the transition temperature is shown in a phase diagram. A rotating image of Fig-\ref{fig:pbd} can be found via the link \url{https://youtu.be/DsAakI0Yjdw}, for better understanding. The solid red dots denote the critical points whereas the compensation points are represented by the hollow blue dots. The phase space can be divided into two different areas of interest. First, we have a ferrimagnetic phase space for which there is no compensation effect at any temperature and on the other hand, there is another ferrimagnetic phase where compensation phenomenon takes place at temperature $T_{comp}$ . First, we like to draw attention to the nature of the diagram. It can be seen that both the $T_{crit}$ and $T_{comp}$ decreases with increasing impurity concentration $C$ as well as increasing antiferromagnetic coupling $J_{AB}$. In Fig-\ref{fig:pbd} when $J_{AB}/J_{BB}=-0.1$, we have strong compensation such that the separation between surfaces is wide, whereas $J_{AB}/J_{BB}=-1.0$, shows no compensation effect and the two surfaces merge together closing the gap. The merging of the gap is clearly visible in Fig-\ref{fig:pbd}d, indicates that any strong value of $J_{AB}$ rules out the presence of the compensation effect. This can be understood as follows:\\
\newpage

For compensation, one needs $|M_2|=|M_1|+|M_3|$ condition to be satisfied i.e. the B-plane sublattice magnetization should be enough to cancel out both the A-plane magnetization. Now suppose the system has no antiferromagnetic strength, $J_{AB}=0$ then each of the layers will behave as a single ferromagnetic system and will have different critical temperatures with no compensation effect. This is obvious as all the three sublattice magnetizations, $M_\sigma$ will either saturate to +1 or -1 for $T<T_c$. Once the $J_{AB}$ coupling is turned on, all the three layer starts to acts as a single system and have a unique critical temperature. Now if the antiferromagnetic strength is more compared to the ferromagnetic strength, the B-plane will get "frozen" at either "+1" state or "-1" state because both the A-layers will try to align the spins of B-layer anti-parallel with respect to each other. This will result in a trilayer ABA with its B-plane sublattice magnetization almost equal to A-plane. The total magnetization, $M_\sigma$ will not be zero for $T<T_{crit}$ and will have a non-zero value which is equal to one of its A-plane. Therefore we need a non-zero $J_{AB}$ in the trilayer for it to act as a single system, also a strong value of   $J_{AB}$ compared to $J_{BB}$ is not desired as the total magnetization never reaches zero for $T<T_c$ and no compensation effect is observed.\\
We see in Fig-\ref{fig:pbd} as the impurity concentration $C$ increases ranging from $0-0.26$ both $T_{crit}$ and $T_{comp}$ decreases. We also report a trend that the slope of the curve decreases faster as the impurity concentration increases ($C$) which is evident in the phase diagram. In order to understand the effect of impurity on $T_{crit}$ and $T_{comp}$ the smallest unit of a square lattice can be imagined where $T\rightarrow 0$ so that the most of the spins oriented in a particular direction. If all the 4 spins of the unit are up we need to supply energy as in the form of $k_BT$ to flip the spins i.e. to break the order and drive the system into the paramagnetic state. But in the case of impurity, we have less active spin sites per unit volume in the system than the pure case. As a result, less amount of energy is needed to flip the spins. Therefore dilution causes the magnetization curve to converge quickly to zero and thus yielding a low value of $T_{crit}$ and $T_{comp}$ for higher impurity concentration. 
\section{Conclusion}
We conducted a study on a site diluted trilayer spin-1/2 ferrimagnetic system. The trilayer is made up of two different layers of theoretical atoms A \& B stacked in a particular fashion like A-B-A. The interaction between a similar atom (intraplanar) is ferromagnetic whereas the cross atom interaction (interplanar) is antiferromagnetic. We write down the Hamiltonian with the help of the Ising Hamiltonian. The system at first is randomly distributed with an equal number of up ($s=+1$) and down ($s=-1$) spins. Later we introduced $s=0$ spins randomly in every layer which mimicked the non-magnetic impurity. The position of the impure sites is constant over time as the impurity is quenched in nature. Our goal is to find the effects of interaction strength and impurity concentration on $T_{crit}$ and $T_{comp}$ and hence obtain the phase diagram along with the lattice morphology. We used Monte-Carlo method(MC) to tackle the problem as it takes fluctuations into account, unlike MF approximation and is also simple yet powerful enough to give accurate results.\\
Initially, we started investigating the lattice morphologies at different situations which is not usually seen in these type of studies. Study of the morphology reveals that the $T_{crit}$ and $T_{comp}$ are two fundamentally different points and their lattice morphologies are different as well. The formation of the asymmetric spin cluster at $T_{comp}$ is highlighted as it helps compensation phenomenon to take place. We also provide the layered magnetizations in each case and observe the vanishing magnetizations at $T_{crit}$ where at $T_{comp}$ sublattice magnetizations remains non zero. Next, we present how impurity affects the thermal and magnetic behaviour. We plotted quantities such as total magnetization (Fig-\ref{fig:mag}), susceptibility (Fig-\ref{fig:suscep}) and specific heat (Fig-\ref{fig:cv}) for dilution $C=0.0, 0.09, 0.18, 0.26$. This shows that the magnetization curve converges quickly as the impurity increases and so $T_{crit}$ and $T_{comp}$ decreases. Finally, we report the phase diagram for a selected range of Hamiltonian parameters $J_{AB}$ and impurity concentration $C$. We obtain the values of $T_{crit}$ and $T_{comp}$ for  $C=0-0.26$ while $J_{AB}/J_{BB}$ is ranging from $-0.1$ to $-1.0$. Our conclusion is as follows: (I)$T_{crit}$ and $T_{comp}$ decreases as $J_{AB}/J_{BB}$ and $C$ increases, (II)phase diagram is divided into two regions, (III)one region is void of any compensation effect at any temperature and (IV)the other one shows the compensation effect for selected values of $J_{AB}/J_{BB}$. The boundary at which the compensation phenomenon vanishes is around $J_{AB}/J_{BB}=-0.9$. Beyond this point, we only have the critical point and do not see any compensation effect.\\
{\it Our main observation is that the non-magnetic impurities reduce both the critical temperature ($T_{crit}$) and the compensation temperature ($T_{comp}$)}. This is particularly useful in the case of magnetic cooling. One of the particular methods of magnetic cooling is adiabatic demagnetization (AD) exploits the paramagnetic properties of some material. It involves the process of sequentially magnetizing and demagnetizing of a paramagnetic substance under the adiabatic condition to reach very low temperature such as 1K or even colder. But if somehow the initial Neel temperature can be lowered, with the same number of steps, we can achieve even lower temperature. One of the popular substance that is used in magnetic cooling is gadolinium (Gd) and its alloys. But being a rare-earth material it is quite expensive. Therefore, to find alternative materials, especially without rare-earth metal, is of great interest \cite{shen,tanabe,fujita}. The layered magnetic materials with properties like compensation phenomenon and being economically cheap compared to other rare-earth metals is a strong candidate for the magnetocaloric material. Thus, the study of magnetic layered systems in terms of MCE is gaining attention \cite{szalowski,franco,florez}. Moreover, as we have shown in our work, the fact that Neel temperatures of a trilayer magnetic material can be lowered using the parameters like impurity and interaction strength may be helpful to achieve even lower temperature using AD. Also, the evaluation of $T_{comp}$ and $T_{crit}$ for different Hamiltonian parameter and impurity concentration along with the investigation of regions where compensation takes place is an invaluable knowledge to the experimentalists. For the future, it would be interesting to study the compensation in the system having a higher value of spin.
\label{conc}

\vspace{0.5cm}

{\bf Acknowledgments:} The authors like to thank Soham Chandra for helpful discussions. SS thanks Pratyusava Baral for careful reading of the manuscript and feedback. MA acknowledges the financial support provided by Presidency University through FRPDF grant.

\end{document}